\RequirePackage{fix-cm}

\documentclass{article}       

\usepackage{graphicx}
\usepackage[caption=false]{subfig} 
\begin{document}

\title{A Big Data Analyzer for Large Trace Logs%\thanks{Grants or other notes
%about the article that should go on the front page should be
%placed here. General acknowledgments should be placed at the end of the article.}
}
%\subtitle{}

%\titlerunning{Short form of title}        % if too long for running head

\author{Alkida Balliu$^1$, Dennis Olivetti$^1$, Ozalp Babaoglu$^2$,\\
        Moreno Marzolla$^2$ and Alina S\^irbu$^2$\\
        \\
        1. Gran Sasso Science Institute (GSSI), L'Aquila, Italy\\
	2. Department of Computer Science and Engineering, \\University of Bologna, Italy
}

%\authorrunning{Balliu et al.} % if too long for running head

\date{}

\maketitle

\begin{abstract}
Current generation of Internet-based services are typically hosted on large data centers that take the form of warehouse-size structures housing tens of thousands of servers. Continued availability of a modern data center is the result of a complex orchestration among many internal and external actors including computing hardware, multiple layers of intricate software, networking and storage devices, electrical power and cooling plants.  During the course of their operation, many of these components produce large amounts of data in the form of event and error logs that are essential not only for identifying and resolving problems but also for improving data center efficiency and management. Most of these activities would benefit significantly from data analytics techniques to exploit hidden statistical patterns and correlations that may be present in the data. The sheer volume of data to be analyzed makes uncovering these correlations and patterns a challenging task. This paper presents \emph{BiDAl}, a prototype Java tool for log-data analysis that incorporates several Big Data technologies in order to simplify the task of extracting information from data traces produced by large clusters and server farms. \emph{BiDAl} provides the user with several analysis languages (SQL, R and Hadoop MapReduce) and storage backends (HDFS and SQLite) that can be freely mixed and matched so that a custom tool for a specific task can be easily constructed. \emph{BiDAl} has a modular architecture so that it can be extended with other backends and analysis languages in the future. In this paper we present the design of \emph{BiDAl} and describe our experience using it to analyze publicly-available traces from Google data clusters, with the goal of building a realistic model of a complex data center.
%\keywords{big data \and log analysis \and workload characterization \and Google cluster trace \and model \and simulation}
% \PACS{PACS code1 \and PACS code2 \and more}
 %\subclass{MSC 68N01 \and MSC 68P20 \and MSC 68U20}
\end{abstract}

%%%%%%%%%%%%%%%%%%%%%%%%%%%%%%%%%%%%%%%%%
%%%%%%%%%%%%%%%%%%%%%%%%%%%%%%%%%%%%%%%%%
%%%%%%%%%%%%%%%%%%%%%%%%%%%%%%%%%%%%%%%%%
\section{Introduction}
\label{intro}
Large data centers are the engines of the Internet that run a vast majority of modern Internet-based services such as cloud computing, social networking, online storage and media-sharing services. A modern data center contains tens of thousands of servers and other components (e.g., networking equipment, power distribution, air conditioning) that may interact in subtle and unintended ways, making management of the global infrastructure a nontrivial task. Failures are extremely costly both for data center operators and their customers, since the services provided by these huge infrastructures have become vital to society in general.  In this light, monitoring and managing large data centers to keep them running correctly and continuously become critical tasks. 

The amount of log data produced by modern data centers is growing steadily, making log management itself technically challenging. For instance, a 2010 Facebook study reports 60 Terabytes of log data being produced by its data centers each day~\cite{Thusoo2010}.  For live monitoring of its systems and analyzing their log data, Facebook has developed a dedicated software tool called Scuba~\cite{Abraham2013} that uses a large in-memory database running on hundreds of servers with 144GB of RAM each. This infrastructure needs to be upgraded every few weeks to keep up with the increasing computational power and storage requirements that Scuba generates.  

Making sense of these huge data streams is a task that continues to rely heavily on human judgement, and is therefore error-prone, time-consuming and potentially inefficient. Log analysis falls within the class of Big Data applications: the data sets are so large that conventional storage and analysis techniques are not appropriate to process them.  There is a real need to develop novel tools and techniques for analyzing logs, possibly incorporating data analytics to uncover hidden patterns and correlations that can help system administrators avoid critical states, or to identify the root cause of failures or performance problems. The ``holy grail'' of system management is to render data centers fully autonomic; ideally, the system should be capable of analyzing its state and use this information to identify performance or reliability problems and correct them or alert system managers directing them to the root causes of the problem. Even better, the system should be capable of anticipating situations that may lead to performance problems or failures, allowing for proactive countermeasures to be put in place in order to steer the system away from undesirable states towards desired operational states. These challenging goals are still far from being realized~\cite{Salfner2010}.

Numerous studies have analyzed trace data from a variety of sources for different purposes (see the related work in Section~\ref{sec:discussion}), but typically without relying on an integrated software framework developed specifically for log analysis~\cite{Chen2012,Liu2012,Reiss2012}. Reasons for this are several fold: first, the amount, content and structure of logs are often system- and application-specific, requiring ad-hoc solutions that are difficult to port to other contexts. Furthermore, log trace data originating from commercial services are highly sensitive and need to be kept strictly confidential. All these facts lead to fragmentation of analysis frameworks and difficulty in porting them to traces from other sources. One isolated example of analysis framework is the Failure Trace Archive Toolkit~\cite{Javadi2013}, limited however to failure traces. Lack of a more general framework for log data analysis results in time being wasted ``reinventing the wheel"  -- developing software for parsing, interpreting and analyzing the data, repeatedly for each new trace~\cite{Javadi2013}. 

As a first step towards realizing the above goals, we present \emph{BiDAl} (Big Data Analyzer), a prototype software tool implementing a general framework for statistical analysis of very large trace data sets. \emph{BiDAl} is built around two main components: a storage backend and an analysis framework for data processing and reduction. Currently, \emph{BiDAl} supports HDFS and SQlite as storage backends, and SQL, R, and Hadoop MapReduce as analysis frameworks. However, \emph{BiDAl} is extensible so that additional backends and analysis frameworks can be easily added, and multiple types can coexist and be used at the same time. After describing the architecture of \emph{BiDAl}, we illustrate how it has been used to analyze publicly-available Google cluster trace data~\cite{googleData} in order to extract parameters of a cluster model which we have implemented. Both the \emph{BiDAl} prototype and the model are  publicly available through a GNU General Public License (GPL)~\cite{bidalCode}.

The contributions of this work are several fold. First, we present \emph{BiDAl} and describe its architecture incorporating several Big Data technologies that facilitate efficient processing of large datasets for data analytics. Then, we describe an application scenario where we use \emph{BiDAl} to extract workload parameters from Google cluster traces. We introduce a model of the Google cluster which allows for simulation of the Google system. Depending on the input to the model, several types of simulations can be performed. Using the exact workload from the Google trace as input, our model is able to faithfully reproduce many of the behaviors that are observed in the traces. By providing the model with distributions of the various parameters that are obtained using \emph{BiDAl} more general workloads can also be simulated; in this scenario, simulation results show that our model is able to approximate average behavior, although variability is lower than in the real counterpart. 
%%AS
%%We could use the next sentence to cite the workshop paper.
%Results have been partially presented in an earlier publication \cite{balliu2014}, while here we provide a complete analysis.

This paper is organized as follows. In Section~\ref{sec:bidal} we provide a high level overview of the framework followed by a detailed description of its components. In Section~\ref{sec:sim} we apply the framework to characterize the workload from a public Google cluster trace, and use this information to build a model of the Google cluster and perform simulations. In Section~\ref{sec:discussion} we discuss related work, and conclude with new directions for future research in Section~\ref{sec:conclusion}.

%%%%%%%%%%%%%%%%%%%%%%%%%%%%%%%%%%%%%%%%%
%%%%%%%%%%%%%%%%%%%%%%%%%%%%%%%%%%%%%%%%%
%%%%%%%%%%%%%%%%%%%%%%%%%%%%%%%%%%%%%%%%%
\section{The Big Data Analyzer (\emph{BiDAl}) prototype}
\label{sec:bidal}

%%%%%%%%%%%%%%%%%%%%%%%%%%%%%%%%%%%%%%%%%
%%%%%%%%%%%%%%%%%%%%%%%%%%%%%%%%%%%%%%%%%
%%%%%%%%%%%%%%%%%%%%%%%%%%%%%%%%%%%%%%%%%
\subsection{General overview}

\begin{figure}[ht]
\centering
  \includegraphics[width=8cm]{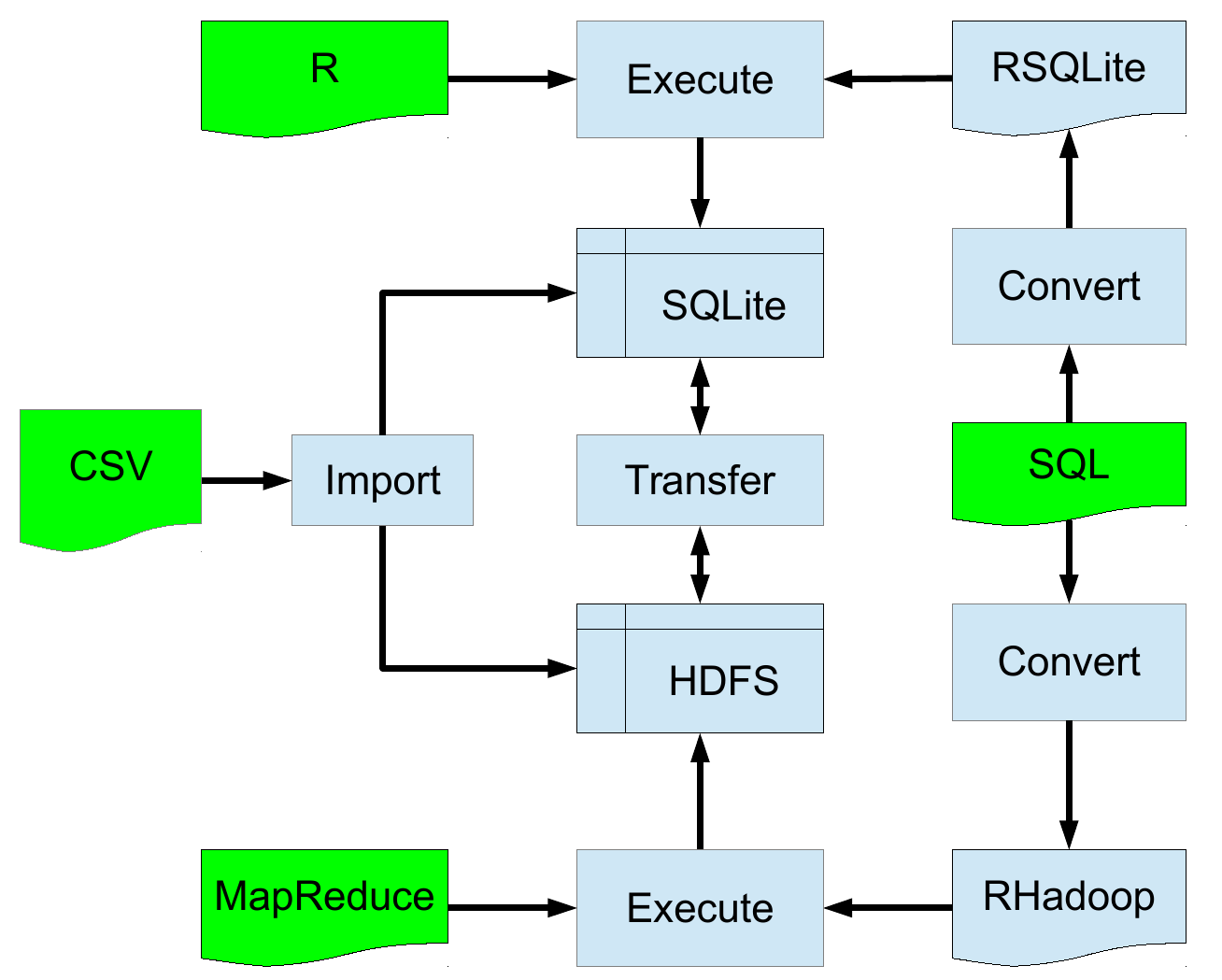}
\caption{Data flow in \emph{BiDAl}. Raw data in CSV format is imported into the selected storage backend, and can be selected or aggregated into new tables using SQL queries. Commands using R or MapReduce can then be applied both to the original imported data and to the derived tables. Data can be automatically and transparently moved between the storage backends.}
\label{fig:data_flow}       
\end{figure}

The typical \emph{BiDAl} workflow consists of three steps: instantiation of a storage backend (or opening an existing one), data selection and aggregation, and data analysis; Figure \ref{fig:data_flow} shows the overall data flow within \emph{BiDAl}. 

For storage creation, \emph{BiDAl} is designed to import CSV files (Comma Separated Values, the typical format for trace data) into an SQLite database or to a Hadoop File System (HDFS) storage, depending on the user?s preference; HDFS is the preferred choice for handling large amounts of data using the Hadoop framework. Except for the CSV format, no other restrictions on the data type exist, so the platform can be easily used for data from various sources, as long as they can be viewed as CSV tables. Even though the storages currently implemented are based on the the concept of tables (stored in a relational database by SQLite and CSV files by Hadoop), other storage types can be supported by \emph{BiDAl}. Indeed, Hadoop supports HBase, a non-relational database that works with $<$key,value$>$ pairs. Since Hadoop is already supported by \emph{BiDAl}, a new storage that works on this type of non-relational databases can be easily added.

Selections and aggregations can be performed through queries expressed using a subset of SQL, for example to create new tables or to filter existing data. SQL queries are automatically translated into the query language supported by the underlying storage system (RSQLite or RHadoop). At the moment, the supported statements in the SQL subset are SELECT, FROM, WHERE and GROUP BY.  Queries executed on the SQL storage do not require any processing, since the backend (SQlite) already supports a larger subset of SQL. For the Hadoop backend, GROUP BY queries are mapped to MapReduce operations. The Map function implements the GROUP BY part of the query, while the Reduce function deals with the WHERE and SELECT clauses.

\emph{BiDAl} can perform statistical data analysis using both R~\cite{r} and Hadoop MapReduce~\cite{hadoop,mapreduce} by offering a set of predefined commands. Commands implemented in R are typically applied to the SQLite storage, while those in MapReduce to the Hadoop storage. However, the system allows mixed execution of both types of commands regardless of the storage used, being able to switch between backends (by exporting data) transparent to the user. For instance, after a MapReduce command, it is possible to analyze the outcome using commands implemented in R; in this case, the software automatically exports the result obtained from the MapReduce step, and imports it to the SQLite storage where the analysis can continue using commands implemented in R. This is particularly useful for handling large datasets, since the volume of data can be reduced by applying a first processing step with Hadoop/MapReduce, and then using R to complete the analysis on the resulting (smaller) dataset. The drawback is that the same data may end up being duplicated into different storage types so, depending on the size of the dataset, additional storage space will be consumed. However, this does not generate consistency issues, since log data does not change once it is recorded.

%%%%%%%%%%%%%%%%%%%%%%%%%%%%%%%%%%%%%%%%%
%%%%%%%%%%%%%%%%%%%%%%%%%%%%%%%%%%%%%%%%%
%%%%%%%%%%%%%%%%%%%%%%%%%%%%%%%%%%%%%%%%%
\subsection{Design}

\begin{figure}
\centering
  \includegraphics[width=12cm]{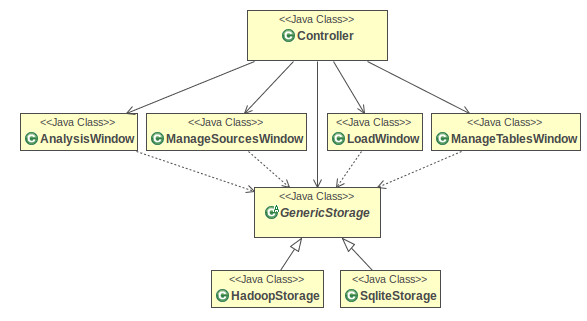}
\caption{UML diagram of \emph{BiDAl} classes. This shows the modular structure where the storage is separated from the user interface, facilitating addition of new types of storage backends. }
\label{fig:uml}       
\end{figure}

\emph{BiDAl} is a modular application designed for extensibility and ease of use. It is written in Java, to facilitate portability across different Operating Systems, and uses a Graphical User Interface (GUI) based on the standard Model-View-Controller (MVC) architectural pattern~\cite{gamma1994}. The View provides a Swing GUI, the Model manages different types of storage backends, and the Controller handles the interaction between the two. Figure~\ref{fig:uml} outlines the architecture using the UML class diagram. 

The Controller class connects the GUI with the other components of the software. The Controller implements the Singleton pattern, with the one instance accessible from any part of the code. The interface to the different storage backends is given by the GenericStorage class, that has to be further specialized by any concrete backend developed. In our case, the two existing concrete storage backends are represented by the SqliteStorage class to support SQLite, and the HadoopStorage class, to support HDFS. Neither the Controller nor the GUI elements communicate directly with the concrete storage backends, but only with the abstract class GenericStorage. This simplifies the implementation of new backends without the need to change the Controller or GUI implementations.

\begin{figure}
\centering
  \includegraphics[width=12cm]{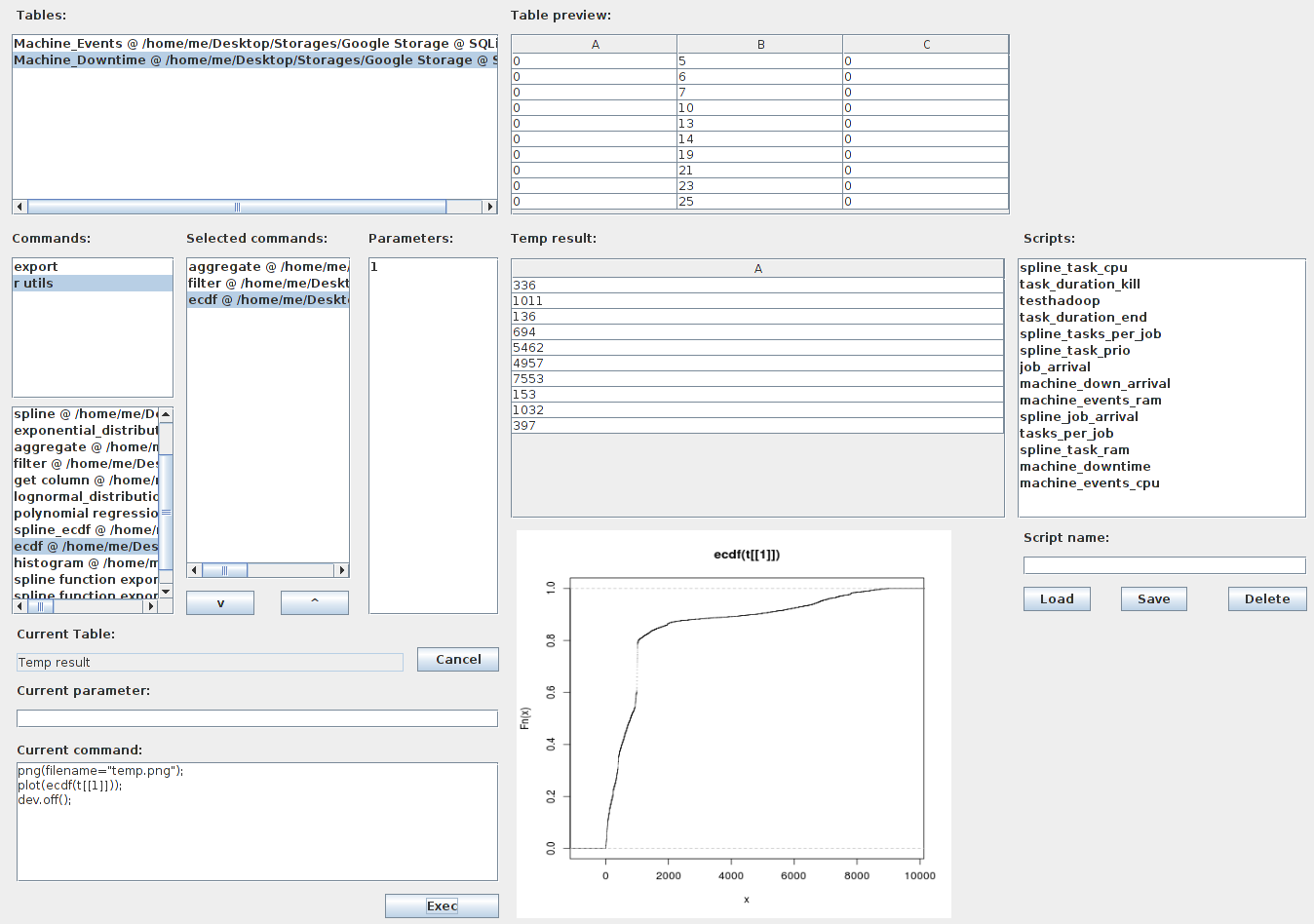}
\caption{Screenshot of the BiDAl analysis console. In the upper-left corner, we see the list of available tables in the current storage backend (SQLite in this case), and the list of available commands (implemented in R).  The results of running the selected command (ecdf) on the selected table (Machine\_Downtime) are shown in the plot at the bottom. The command implementation can be edited in the lower-left panel. New commands can be saved, with a list of existing custom commands displayed in the ``Scripts" panel to the right.}
\label{fig:gui}       
\end{figure}

The user can inspect and modify the data storage using a subset of SQL; the SqliteStorage and HadoopStorage classes use the open source SQL parser Akiban to convert the queries inserted by users into SQL trees that are further mapped to the native language (RSQLite or RHadoop) using the Visitor pattern. The HadoopStorage uses also a Bashexecuter that allows to load files on the HDFS using bash shell commands. A new storage class can be implemented by providing a suitable specialization of the GenericStorage class, including the mapping of the SQL tree to specific commands understood by the backend.

Although the SQL parser supports the full SQL language, the developer must define a mapping of the SQL tree into the language supported by the underlying storage; this often limits the number of SQL statements that can be supported due to the difficulty of realizing such a mapping.

%%%%%%%%%%%%%%%%%%%%%%%%%%%%%%%%%%%%%%%%%
%%%%%%%%%%%%%%%%%%%%%%%%%%%%%%%%%%%%%%%%%
%%%%%%%%%%%%%%%%%%%%%%%%%%%%%%%%%%%%%%%%%
\subsection{Using R with \emph{BiDAl} }

\emph{BiDAl}  provides a list of predefined commands, implemented in R, that can be selected by the user from a graphical interface (see Figure~\ref{fig:gui} for a screenshot and Table \ref{tab:comm} for a partial list of the available commands). When a command is selected, an input box appears asking the user to provide the parameters needed by that specific command. Additionally, a text box (bottom-left corner of Figure \ref{fig:gui}) allows the user to modify on the fly the R code to be executed.

\begin{table}
\begin{tabular}{|p{3.5cm} p{8cm}|}
\hline%\noalign{\smallskip}
{\bf \emph{BiDAl} command} & {\bf Description}   \\
%\noalign{\smallskip}
\hline\hline
%\noalign{\smallskip}
get\_column&Selects a column.\\
\hline
apply\_1Col&Applies the desired R function to each element of a column.\\
\hline
aggregate&Takes as input a column to group by; among all rows selects the ones that satisfies the specified condition; the result obtained is specified from the R function given to the third parameter.\\
\hline
difference\_between\_rows&Calculates the differences between consecutive rows.\\
\hline
filter&Filters the data after the specified condition.\\
\hline
exponential\_distribution&Plots the fit of the exponential distribution to the data.\\
\hline
lognormal\_distribution&Plots the fit of the lognormal distribution to the data.\\
\hline
polynomial\_regression&Plots the fit of the n-grade polynomial regression to the data in the specified column.\\
\hline
ecdf&Plots the cumulative distribution function of the data in the specified column.\\
\hline
spline&Divides the data in the specified column in n intervals and for each range plots spline functions. Also allows to show a part of the plot or all of it.\\
\hline
log\_histogram&Plots the histogram of the data in the specified column, using a logarithmic y-axis. \\
%\noalign{\smallskip}
\hline
\end{tabular}
\caption{A partial list of \emph{BiDAl} commands implemented in R}
\label{tab:comm}     
\end{table}

All commands are defined in an external text file. New operations can therefore be added quite easily by simply including them in the file.

%%%%%%%%%%%%%%%%%%%%%%%%%%%%%%%%%%%%%%%%%
%%%%%%%%%%%%%%%%%%%%%%%%%%%%%%%%%%%%%%%%%
%%%%%%%%%%%%%%%%%%%%%%%%%%%%%%%%%%%%%%%%%
\subsection{Using Hadoop/MapReduce with \emph{BiDAl} }

\emph{BiDAl}  allows computations to be distributed across many machines through the Hadoop/MapReduce abstractions. The user can access any of the builtin commands implemented in RHadoop, or create new ones. Usually, the Mapper and Reducer are implemented in Java, generating files that need to be compiled and then executed. However, \emph{BiDAl}  abstracts from this approach by using the RHadoop library which handles MapReduce job submission and permits to interact with Hadoop's file system HDFS using R. This allows for reuse of the \emph{BiDAl}  R engine for the Hadoop backend. Once the dataset of interest has been chosen, the user can execute the Map and Reduce commands implemented in RHadoop or create new ones. Again, the commands and corresponding RHadoop code are saved in an external text file, using the same format described above, so the creation of new commands does not require any modification to \emph{BiDAl}  itself. At the moment, one Map command is implemented in \emph{BiDAl} , which groups the data by the values of a column. A Reduce command is also available, which counts the elements of each group. Other commands can be added by the user, similar to those implemented in R.

%%%%%%%%%%%%%%%%%%%%%%%%%%%%%%%%%%%%%%%%%
%%%%%%%%%%%%%%%%%%%%%%%%%%%%%%%%%%%%%%%%%
%%%%%%%%%%%%%%%%%%%%%%%%%%%%%%%%%%%%%%%%%
\section{Case study}
\label{sec:sim}

The development of \emph{BiDAl} was motivated by the need to process large data from cluster traces, such as those publicly released by Google~\cite{googleData}. Our goal was to extract workload parameters from the traces in order to instantiate a model of the compute cluster capable of reproducing the most important features observed in the real data. The model, then, could be used to perform ``what-if analyses'' by simulating different scenarios where the workload parameters are different, or several types of faults are injected into the system.

In this section we first present the structure of the model, then describe the use of \emph{BiDAl}  for analyzing the Google traces and extracting parameters for the model.

%%%%%%%%%%%%%%%%%%%%%%%%%%%%%%%%%%%%%%%%%
%%%%%%%%%%%%%%%%%%%%%%%%%%%%%%%%%%%%%%%%%
%%%%%%%%%%%%%%%%%%%%%%%%%%%%%%%%%%%%%%%%%
\subsection{Modeling the Google compute cluster}
\label{sec:model}

We built a model of the Google compute cluster corresponding to that from which the traces were obtained. According to available information, the Google cluster is basically a large batch system where computational tasks of different types are submitted and executed on a large server pool. Each job may describe constraints for its execution (e.g., a minimum amount of available RAM on the execution host); a scheduler is responsible for extracting jobs from the waiting queue, and dispatching them to a suitable execution host. As can be expected on a large infrastructure, jobs may fail and can be resubmitted; moreover, execution hosts may fail as well and be temporarily removed from the pool, or new hosts can be added. The Google trace contains a list of timestamped events such as job arrival, job completion, activation of a new host and so on; additional (anonymized)  information on job requirements is also provided.

\begin{figure}
\centering
  \includegraphics[width=10cm]{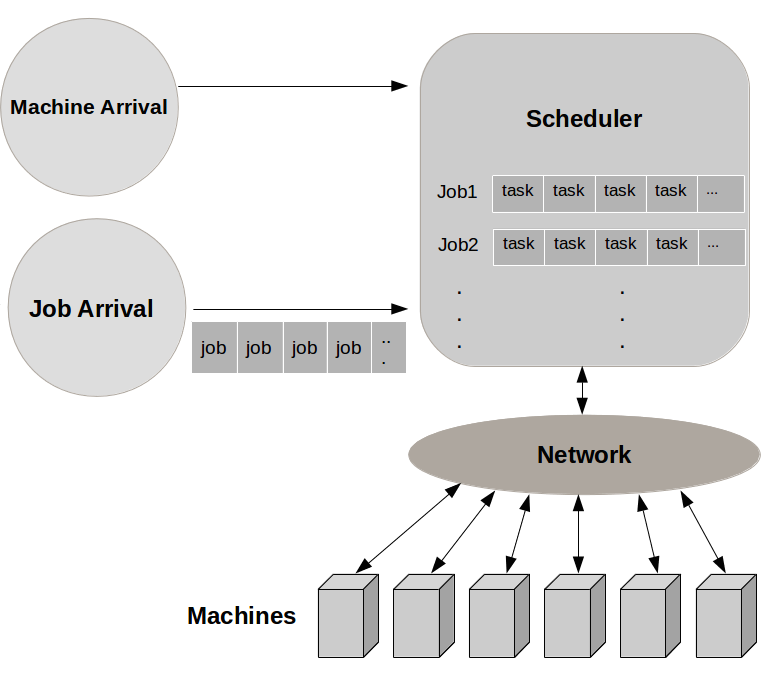}
\caption{Simple model of a Google compute cluster. This includes active entities that exchange messages among each other (Machine Arrival, Job Arrival, Scheduler, Network and Machine), and passive entities that are silent (Jobs and Tasks). The arrows show the flow of information. }
\label{fig:sim}       
\end{figure}

The model, shown in Figure~\ref{fig:sim}, consists of several active and passive interacting entities. The passive entities (i.e., those that do not exchange any message with other entities) are Jobs and Tasks. The active entities are those that send and receive messages: Machine, Machine Arrival, Job Arrival, Scheduler and Network. The model was implemented using C++ and Omnet++~\cite{varga2001}, a discrete-event simulation tool.

A Task represents a process in execution, or ready to be executed. Each task is characterized by its Id, the information regarding the \emph{requested} and \emph{used} resources (CPU and RAM), its priority, duration, termination cause and other information regarding the execution constraints. Note that the total duration of a task, the termination cause and the effective use of resources are \emph{not} used to take decisions (for example on which machine to execute a task). This choice is necessary in order to simulate a real scenario, where one does not know in advance the length, exit code and resource usage of a task. 

A Job is identified by a unique ID, and can terminate either because all of its tasks complete execution, or because it is aborted by the submitting user. Note that, according to the functioning of the Google cluster, tasks from the same job do not necessarily have to be executed at the same time. The Job Arrival entity generates events that signal new jobs being submitted. At each event, a Job entity is created and sent to the scheduler.

The Machine entity represents an execution node in the compute cluster. Each machine is characterized by an Id and its maximum amount of free resources. Machine Arrival is the entity in charge of managing all machines, and generates events related to addition or removal of machines to the cluster, as well as update events (when the maximal resources of the machine are changed). In all of these cases, the Machine entity is notified of these changes. At regular intervals, each Machine will notify the Scheduler about the free resources owned. Free resources are computed as the difference between the total resources and those used by all tasks running on that machine. The resources used by each task are considered to be equal to the \emph{requested} amount for the first 5 minutes, then equal to the \emph{average used} amount extracted from the traces. This strategy was adopted after careful analysis of the traces, without knowing any details about the system producing the traces (Borg). Recent publication of Borg details~\cite{verma2015} confirms that our scheduling strategy is very similar to the real system. In the Google cluster, the requested resources are initially reserved (just like in our case), and at five minute intervals the reservation is adjusted based on the real usage and a safety margin through a so-called \emph{resource reclamation} mechanism.

The Scheduler implements a simple job scheduling mechanism. Each time a job is created by the Job Arrival entity, the scheduler inserts its tasks in the ready queue. For each task, the scheduler examines which execution nodes (if any) match the task constraints; the task is eventually sent to a suitable execution node. Due to the fact that the Scheduler does not know in real time the exact amount of free resources for all machines, it may happen that it sends a task to a machine that can not host it. In this case, the machine selects a task to interrupt (evict) and sends it back to the scheduler. Similar to the scheduling policies implemented by the Google cluster, we allow a task with higher priority to evict a running task with lower priority. The evicted task will be added back to the ready queue.

When the machine starts the execution of a task, it generates a future event: the termination event, based on the duration generated/read from the input data. The system does not differentiate between tasks that terminate normally or because they are killed or they fail; the only distinction is for evicted tasks, as explained previously. When the termination event will be handled, the scheduler will be notified by a message. Note that the duration is used only to generate the event, it is not used to make decisions. This is necessary in order to simulate a scenario in which the execution time of a task is not known a priori. In case the task is evicted, this event is deleted and will be recreated when the task will be restarted. 

Finally, the Network entity is responsible for exchanging messages between the other active entities.  In this way, it is possible to use a single gate to communicate with every other entity. Messages include notifications of new jobs arriving, tasks being submitted to a machine, machines reporting their status to the scheduler, etc.  Each message holds 2 different IDs: the sender and the receiver, and the network will be responsible to correctly route the messages by interfacing with the Omnet framework. This scenario reflects the real configuration of Google datacenter where there is a common shared network and the storage area is uniformly accessible from each machine. It was not possible to give a limit to the bandwidth while the latency of the channels is considered to be null. This does not affect the simulation since in Google clusters, the internal network does not seem to be a bottleneck. However it is possible to extend the Network entity in order to implement a latency and a maximal bandwidth between channels.

The Google traces contain information about both \emph{exogenous} and \emph{endogenous} events. Exogenous events are those originating outside the system, such as jobs and machines arrivals or job characteristics; endogenous events are those originating inside the system, such as jobs starting/finishing execution, failure events and similar.

%%AS
%%%in the next sentence we have to see what workload refers to, Moreno will clarify. Or we can just remove the parenthesis.
%%AS
%%for now I removed the parenthesis. We can add it back when we do the first revision, if by then we can answer the question. I think workload refers to both type of events...
In order to instantiate the model and perform simulations, several parameters concerning endogenous and exogenous events 
%(also called \emph{workload}) 
have to be provided as input. The implementation provides two input options:

\begin{itemize}
\item \emph{Synthetic-trace-driven simulation}: in this mode, the simulator is provided with distributions of the various job characteristics and event probabilities and inter-arrival times. For instance, one can specify distributions for number of tasks per job, required and used resources, priorities, and others. During simulation, these distributions are used by the Job Arrival and Machine Arrival entities to generate new Job entities and machine events, obtaining in this way a synthetic trace. Distributions can be specified in two ways. One is by providing CDFs extracted from real traces. We will demonstrate this case in Section~\ref{sec:synth}, when we will extract distributions from the Google trace using \emph{BiDAl} and we will perform simulations. The second option is to specify in a configuration file known distributions for the parameters. For instance, one can use a Gaussian distribution for resource utilization. Synthetic-trace-driven simulation is useful for exploring the behavior of the Google cluster under arbitrary conditions, e.g., under heavy load or massive failures, that may not occur in the traces; this is an example of ``what-if analysis".

\item \emph{Real-trace-driven simulation}: in this mode, information regarding jobs and machine arrivals is contained in a file that is provided at the beginning of the simulation. This includes all properties of each incoming job as described by a trace, and the exact times when machines are added, removed or updated. The data is used by the Job Arrival and Machine Arrival entities to reproduce exactly the same workload during simulation. Trace-driven simulation is used to validate the model, since we expect the output of the simulation runs to match the Google cluster behavior observed in the traces. In Section~\ref{sec:real} we show results from simulation using the Google traces. 
\end{itemize}

%%%%%%%%%%%%%%%%%%%%%%%%%%%%%%%%%%%%%%%%%
%%%%%%%%%%%%%%%%%%%%%%%%%%%%%%%%%%%%%%%%%
%%%%%%%%%%%%%%%%%%%%%%%%%%%%%%%%%%%%%%%%%
\subsection{Synthetic-trace-driven simulation}
\label{sec:synth}

To generate realistic synthetic traces, we used \emph{BiDAl} to extract distributions from the Google data to characterize the workload of the cluster and other relevant endogenous events. It is worth observing that the traces consist of many large CSV files containing records about job and task events, resources used by tasks, task constraints, and so on. There are more than 2000 files with over 1.3 billion records describing the workload and machine attributes for 12,453 cluster nodes, occupying about 40GB in \emph{compressed} form.

In the following we first describe the distribution obtained, then we show simulation results.

%%%%%%%%%%%%%%%%%%%%%%%%%%%%%%%%%%%%%%%%%
%%%%%%%%%%%%%%%%%%%%%%%%%%%%%%%%%%%%%%%%%
%%%%%%%%%%%%%%%%%%%%%%%%%%%%%%%%%%%%%%%%%
\paragraph{Workload Characterization of the Google Cluster}
~

We extracted the arrival time distribution of each job, the distribution of the number of tasks per job, and the distributions of execution times of different types of tasks (e.g., jobs that successfully completed execution, jobs that are killed by the users, and so on). These distributions are used by the model to generate jobs into the system. Additionally, we analyzed the distribution of machines downtime and of the time instants when servers are added/removed from the pool.
 
\begin{figure}
\centering
\subfloat [RAM requested by tasks. Values are normalized by the maximum RAM available on a single node in the Google cluster.]{ \includegraphics[width=5.5cm]{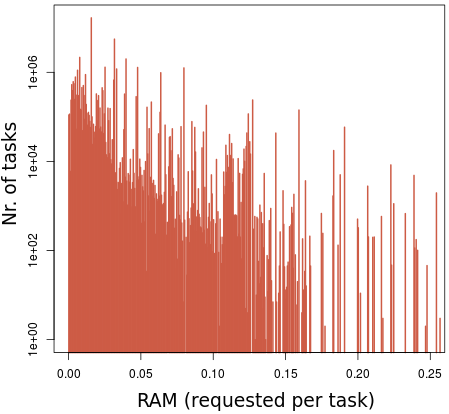} \label{fig:ram} }
 \qquad
   \subfloat[Number of tasks per job]{  \includegraphics[width=5.5cm]{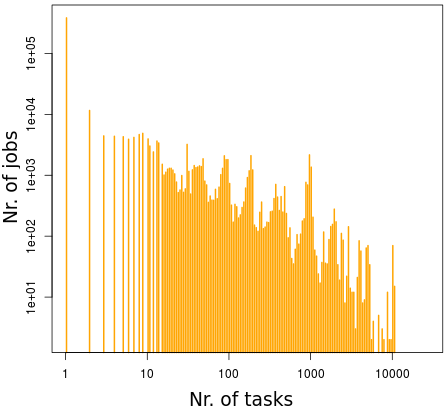}\label{fig:task}} 
\caption{Examples of distributions obtained with BiDAl. These do not appear to follow any known distribution.}      
\end{figure}

Some of the results obtained with \emph{BiDAl} are shown in the following figures (these are the actual plots that were produced by \emph{BiDAl}). Figure~\ref{fig:ram} shows the the amount of RAM requested by tasks, while Figure~\ref{fig:task}  shows the distribution of number of tasks per job.

To generate the graph in Figure~\ref{fig:task}, we first extracted the relevant information from the trace files.  Job and task IDs were required, therefore we generated a new table, called \emph{job\_task\_id}, from the \emph{task\_events.csv} files released by Google~\cite{googleData}. The query generation is automated by \emph{BiDAl} which allows for simple selection of columns using the GUI. Since the DISTINCT clause is not yet implemented in \emph{BiDAl}, we added it manually in the generated query.  The final query used was:

{\footnotesize \begin{verbatim}SELECT DISTINCT V3 AS V1,V4 AS V2 FROM task_events \end{verbatim}}

Here V3 is the \emph{job\_id} column while V4 represents the \emph{task\_id}. On the resulting \emph{job\_task\_id} table, we execute another query to estimate how many tasks each job has, generating a new table called \emph{tasks\_per\_job}:

{\footnotesize\begin{verbatim}SELECT V1 AS V1, COUNT(V2) AS V2 FROM job_task_id GROUP BY V1\end{verbatim}}

Three \emph{BiDAl} commands were used on the  \emph{tasks\_per\_job} table to generate the graph. The first extracts the second column (job id), the second filters out some uninteresting data and the third plots the result. The \emph{BiDAl} commands used are shown in Table~\ref{tab:comm1}.

\begin{table}[h]
\begin{tabular}{|p{2cm}| p{5cm} |p{3cm}|}
\hline%\noalign{\smallskip}
Command & Parameter type &Parameter value   \\
%\noalign{\smallskip}
\hline\hline
%\noalign{\smallskip}
get\_column& column number&2\\ \hline
filter&condition& t[[1]]$<$11000.\\ \hline
log\_histogram&column number, log step, log axis& 1, 0.06, xy\\ 
%\noalign{\smallskip}
\hline
\end{tabular}
\caption{Commands used to generate Figure~\ref{fig:task}}
\label{tab:comm1}     
\end{table}

The analysis was performed on a 2.7 GHz i7 quad core processor with 16GB of RAM and a hard drive with simultaneous read/write speed of 60MB/s. For the example above, importing the data was the most time consuming step, requiring 11 minutes to load 17GB of data into the SQLite storage (the load time is determined for the most part by the disk speed). However, this step is required only once. The first SQL query took about 4 minutes to complete, while the second query and the \emph{BiDAl} commands were almost instantaneous.

\begin{figure}
\centering
  \includegraphics[width=\textwidth]{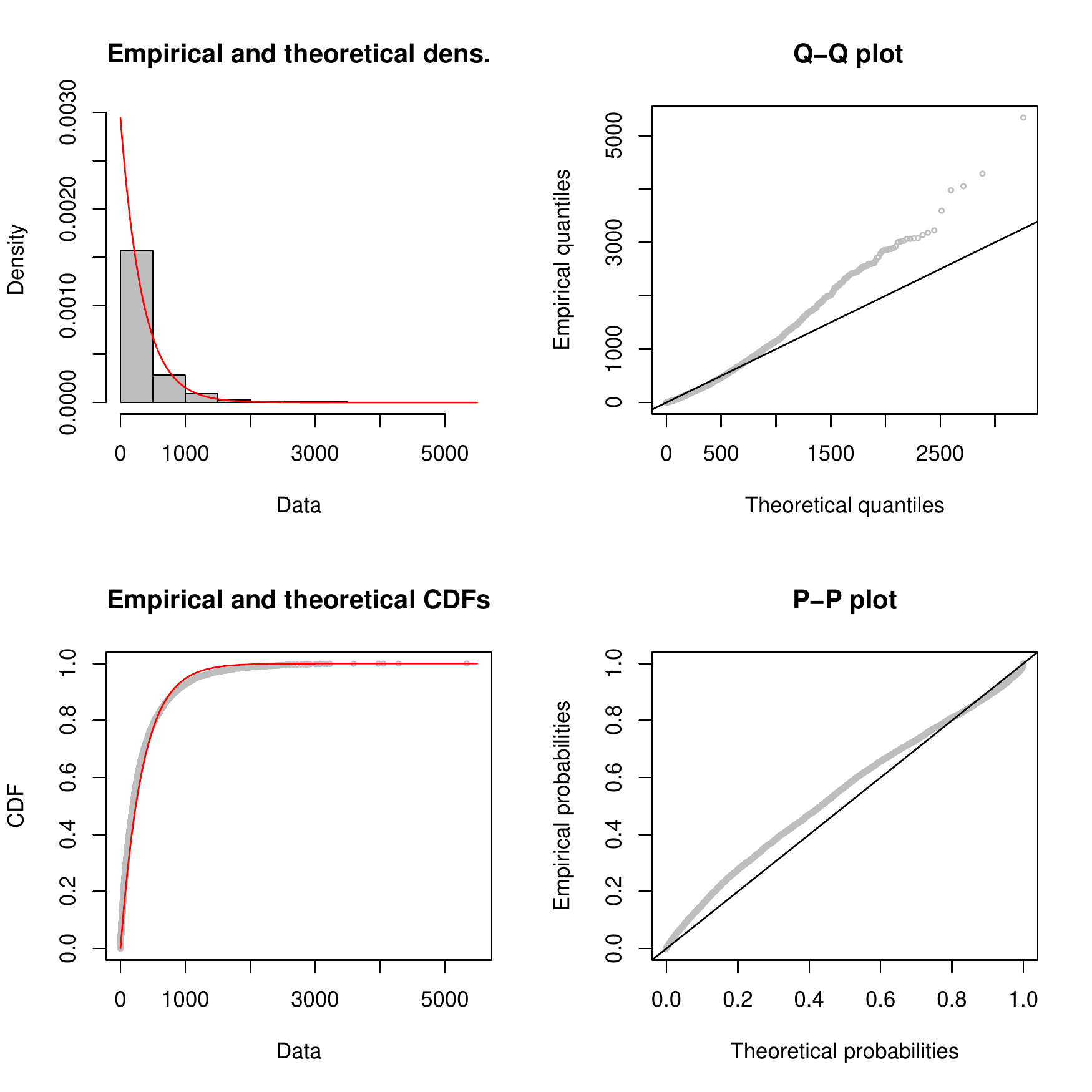}
\caption{Machine update inter-event times, fitted with  an exponential distribution. The left panels show the density and cumulative distribution functions, with the lines representing exponential fitting and the bars/circles showing real data. The right panels show goodness of fit in Q-Q and P-P plots (straight lines show perfect fit).}
\label{fig:time}       
\end{figure}

In Figure~\ref{fig:time} we fit the time between consecutive machine update events (i.e., events that indicate that a machine has changed its list of resources) with an exponential distribution.  We use four standard plots for the goodness of fit: the probability density distribution, the cumulative distribution, the Q-Q (Quantile-Quantile) plot, and P-P (Probability-Probability) plot~\cite{gibbons2011}. The P-P plot displays the values of the cumulative distribution function (CDF) of the data (empirical probabilities) versus the CDF of the fitted exponential distribution (theoretical probabilities). Specifically, for each value $i$ of the inter-event time, the $x$-axis shows the percentage of values in the \emph{theoretical exponential distribution} that fall below $i$ while the $y$-axis shows the percentage of points in the \emph{data} that fall below $i$. If the two values are equal, i.e. the entire plot follows the diagonal, then the fit between the data and theoretical distributions is good. The Q-Q plot, on the other hand, displays the quantiles of the data (empirical quantiles) versus those in the fitted exponential distribution (theoretical quantiles). Again, perfect fit means the Q-Q plot follows the diagonal, i.e. quantiles coincide. All plots show that the observed data is in good agreement with the fitted distribution.

\begin{figure}
\centering
\subfloat [CPU task requirements]{ \includegraphics[width=5.5cm]{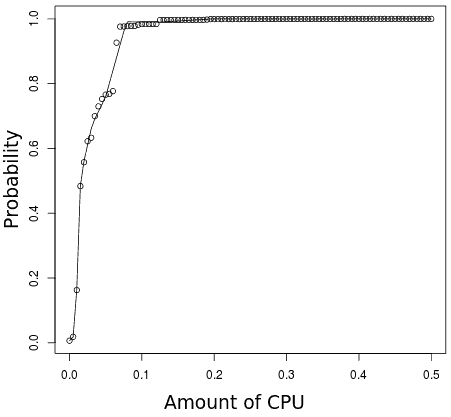} \label{fig:cpu} }
 \qquad
   \subfloat[Machine downtime]{  \includegraphics[width=5.5cm]{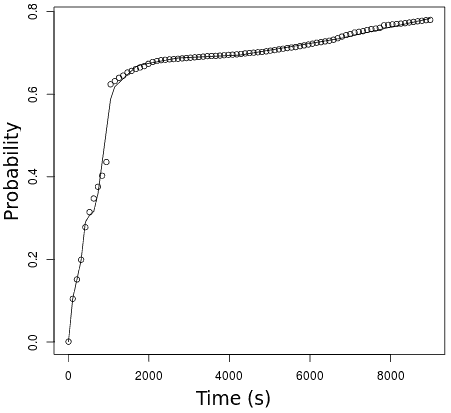}\label{fig:downtime}} 
\caption{Examples of CDFs fitted by sequences of splines, obtained with BiDAl. The circles represent the data, while the lines show the fitted splines. The CDFs are employed to produce synthetic traces to be used as input to our model.}      
\end{figure}

Cumulative distribution functions have also been computed from the data and fitted with sequences of splines, in those cases where the density functions were too noisy to be fitted with a known distribution. For instance, Figure~\ref{fig:cpu} shows the distribution of CPU required by tasks while Figure~\ref{fig:downtime} shows machine downtime, both generated with \emph{BiDAl}. Several other distributions were generated in a similar way to enable simulation of the Google cluster: RAM required by tasks; Task priority; Duration of tasks that end normally; Duration of killed tasks; Tasks per job; Job inter-arrival time; Machine failure inter-arrival time; Machine CPU and RAM.

Once distributions were generated, integration in the model was straightforward since \emph{BiDAl} is able to generate C code related to the different distributions found. In our study, the distributions, hence the C code related to them, represent empirical CDFs.

We extracted several other parameters with \emph{BiDAl} to be used by the model: the probability of submitting tasks with different constraints; the probability that a machine satisfies a constraint; the amount of initial tasks running; the probability of submitting long running tasks (executing from the beginning until the end of the simulation); the amount of RAM available on the machines; the probability that a task terminates normally or is killed.

Jobs constraints were simplified in the synthetic traces compared to real data. For this purpose, we analyzed the traces and studied the influence of the constraints. We calculated the percentage of tasks with constraints and the mean satisfiability $s_{c_i}$ of each constraint $c_i$ as the average fraction of machines that satisfy $c_i$. To simulate the constraint system and assign the same mean satisfiability to each constraint, each machine is associated a numerical value $x$ in an interval $I=[a,b]$. Each constraint $c_i$ is assigned a subinterval $I_{c_i}=[c,d] \subseteq I$ so that $\frac{d-c}{b-a}=s_{c_i}$. A machine satisfies a constraint $c_i$ if $x \in I_{c_i}$. In this way, each constraint is satisfied with the same probability detected from the traces.

%%%%%%%%%%%%%%%%%%%%%%%%%%%%%%%%%%%%%%%%%
%%%%%%%%%%%%%%%%%%%%%%%%%%%%%%%%%%%%%%%%%
%%%%%%%%%%%%%%%%%%%%%%%%%%%%%%%%%%%%%%%%%
\paragraph{Simulation results using synthetic workload}

\begin{figure}
\centering
  \includegraphics[width=\textwidth]{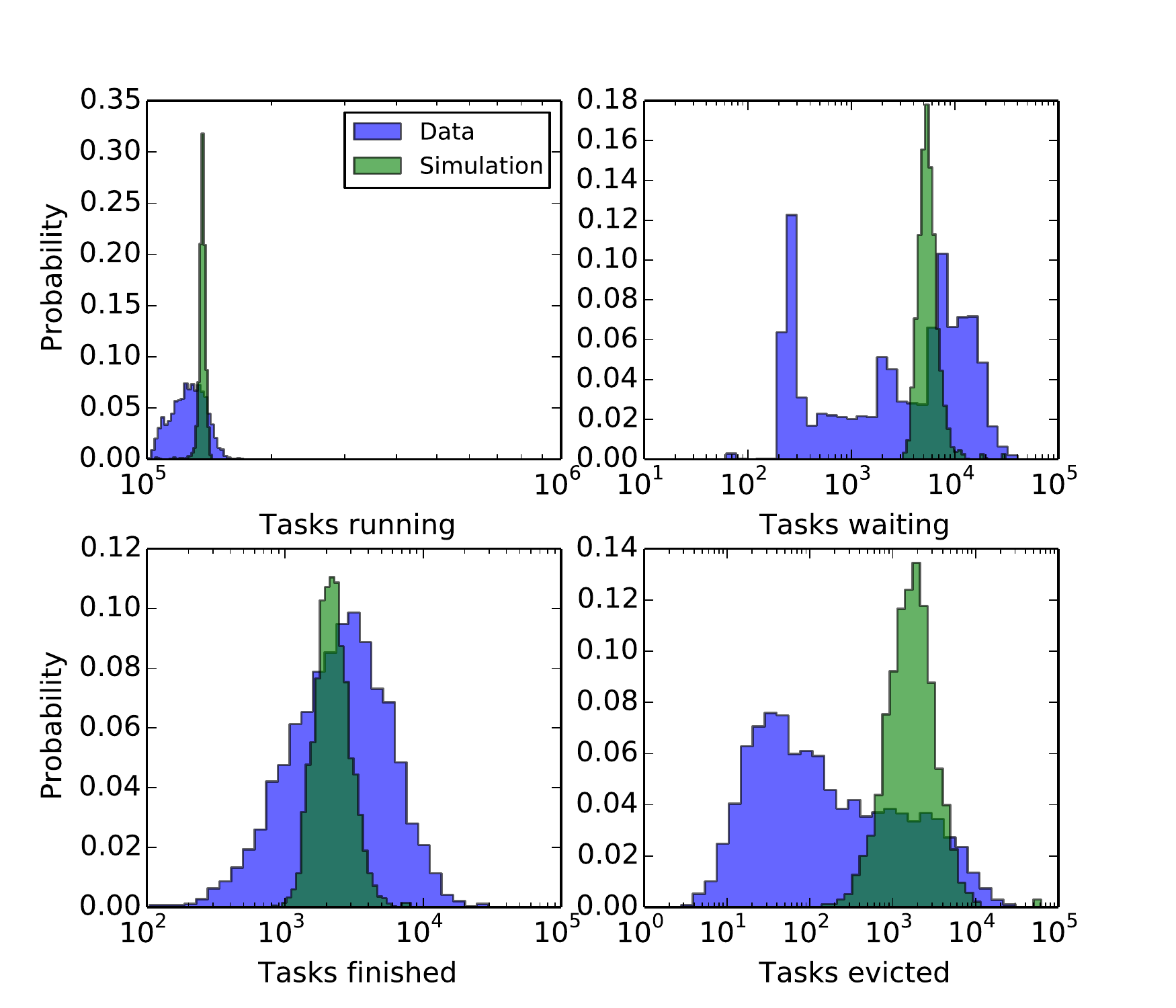}
\caption{Distribution of number of tasks in different categories for each 450s time window, for the \emph{synthetic-trace-driven simulation} compared to the \emph{real data}. The y-axis shows the fraction of time windows with the corresponding number of tasks running, waiting, finished or evicted.  Average behavior is similar between simulation and data for all categories except tasks evicted. However, variability is larger in the data for all cases. }
\label{fig:diff}       
\end{figure}

The parameters and distributions described above have been used to instantiate the model.  We performed ten simulation runs and the results were analyzed in terms of number of running and completed tasks, the length of the ready queue and the number of evicted processes. The distribution of these values, compared to the original data, are shown in Figure~\ref{fig:diff}; Table~\ref{tab:diff} reports the difference between the means of the distributions (real vs. simulated).

\begin{table}[b]
\begin{tabular}{p{1.7cm} p{1.5cm} p{1.5cm} p{1.5cm} p{1.5cm} p{1.5cm}}
\hline%\noalign{\smallskip}
&Running tasks & Ready tasks  &Finished tasks & Evicted tasks& Submitted tasks   \\
%\noalign{\smallskip}
\hline\hline
%\noalign{\smallskip}
Difference & 7\% & 4\% & 29\% & 105\% & 4\%\\ 
%\noalign{\smallskip}
\hline
\end{tabular}
\caption{Differences between means of the distributions from Figure~\ref{fig:diff}. For most measures, averages are very similar. Larger differences are observed for finished and evicted tasks, with our system evicting more and finishing less jobs in each time window, compared to the real system.}
\label{tab:diff}     
\end{table}

The number of tasks in execution, the length of the ready queue and finished tasks are on average similar to the real traces, indicating that the parameters used are fairly good approximations of the real data. However the distributions of these quantities produced by the simulator have a lower variability than those observed in the real data. This is probably due to the fact that resource usage for tasks is averaged over the entire length of the task, rather than being variable in time, as in the real system.

In terms of the number of evicted tasks, differences among average behaviors are much larger. The model tends to evict twice as many tasks as the real system. The mean of the simulation output still falls within a standard deviation from the mean of the real data; however, the simulation never generates low numbers of evicted jobs as are observed in the traces. This can be due, again, to the fact that the simulator is fed with average values for the resource usage and other parameters. Indeed, the same problem is observed, to a smaller extent, also in the real-trace-driven simulation described in the next section. Indeed, resource usage is averaged in real-trace-driven simulation as well.

The number of submitted tasks needs a separate discussion. This metric is different from the other ones because the number of submitted tasks is derived directly from the input distribution, and therefore does not depend on the model; in other words, this is derived from an \emph{input} parameter, rather than the simulation output, so it shows how well \emph{BiDAl} is capable of producing accurate synthetic traces. The number of submitted tasks depends on the distributions of the job inter-arrival time and of the number of tasks per job. 

\begin{figure}
\centering
  \includegraphics[width=0.6\textwidth]{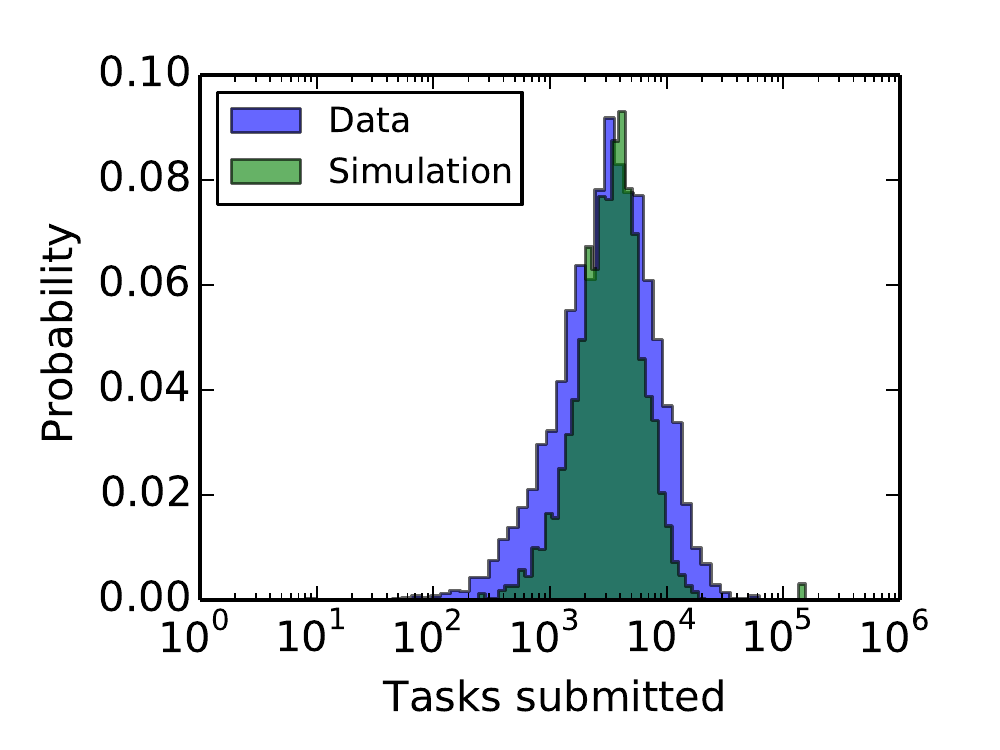}
\caption{Distribution of number of submitted  tasks for the synthetic workload (\emph{Simulation}), compared to the real workload (\emph{Data}). The synthetic workload shows less variation than the real workload. }
\label{fig:diff1}       
\end{figure}

Figure~\ref{fig:diff1} compares the distribution of the number of submitted tasks as seen during simulation and on the real data. The two distributions are very similar; the synthetic trace shows slightly lower variability than the real data, which partly explains why the simulation output has lower variability as well compared to the real data (see Figure~\ref{fig:diff}).

The results indicate that some fine tuning of the model is necessary to get more accurate results. First, the input distributions should better reflect the real data, especially for the arrival rate of tasks. To obtain wider distributions of the number of tasks in the different states, resource usage should be allowed to change over time (as happens in the real data). Furthermore, other system parameters, such as the resource usage limit, should be studied in more detail to get better fits.

%%%%%%%%%%%%%%%%%%%%%%%%%%%%%%%%%%%%%%%%%
%%%%%%%%%%%%%%%%%%%%%%%%%%%%%%%%%%%%%%%%%
%%%%%%%%%%%%%%%%%%%%%%%%%%%%%%%%%%%%%%%%%
\subsection{Real-trace-driven simulation}
\label{sec:real}

In the real-trace-driven simulation we provide the simulation model with the real workload extracted from the traces. The purpose is to validate the model by comparing the simulation results with the real system behavior inferred from the traces.

The Google trace have been collected on a running system; therefore, some jobs were already in the queue, and others were being executed at the beginning of the trace. To properly account for these jobs, we bootstrap the simulation by inserting all the tasks already in execution at the beginning of the trace into the ready queue. These jobs are processed by the scheduler and assigned to machines. At the same time, new Job and Machine events are generated, according to the trace itself. It takes several minutes of wallclock time for the simulation to stabilize and reach a configuration similar to the system state at the beginning of the trace. This phase represents the initial transient and has been removed from the results. The model takes as input the events of the first 40 hours of the original traces, with the first 5 hours considered as part of the initial transient phase.

Running our simulation, we observed that all jobs were scheduled very quickly, with no evicted tasks. However, the Google trace contains many task evictions. The description of the Google data indicates that some machine resources are reserved by the scheduler for itself and for the operating system, so not all resources are available to tasks~\cite{reissgoogle}. This reserved amount is however not specified. We can account for the unknown reserved resources by decreasing the amount of resources available to tasks within the model. We decided to decrease the amount of available memory to a fraction $f_m$ of the total. After several simulations for fine tuning, the value $f_m=0.489$ produced the best fit to the data. The accuracy of our simulation is highly sensitive to this parameter and small variations result in large differences.  For instance, for values slightly different from $0.489$, the number of jobs evicted during simulation is very different from the real traces.  The value obtained for $f_m$ may seem rather large, since it is unlikely that the scheduler reserves half the memory for itself. However, this accounts also for the slight difference in allocating resources in our model compared to the real system. In our case, we reserve exactly the used resources, while the Google cluster, within its resource reclamation mechanism described in Section~\ref{sec:model}, uses a safety margin which is not specified. Our chosen value $f_m=0.489$ includes both the unknown resource reclamation margins and operating system reservation. 
\begin{figure}
\centering
\subfloat [Number of running tasks]{ \includegraphics[width=5.5cm]{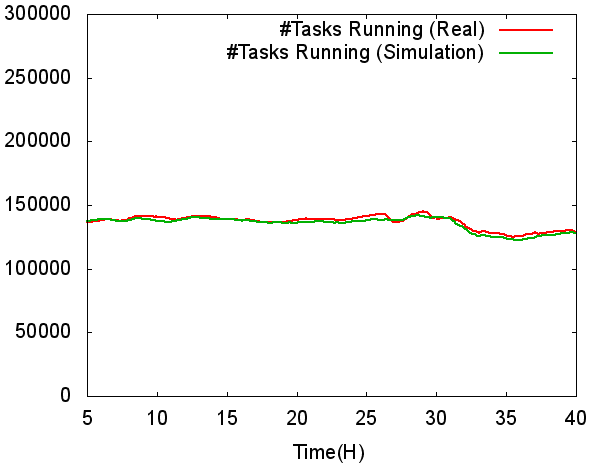} \label{fig:running} }
 \qquad
   \subfloat[Number of tasks completed]{  \includegraphics[width=5.5cm]{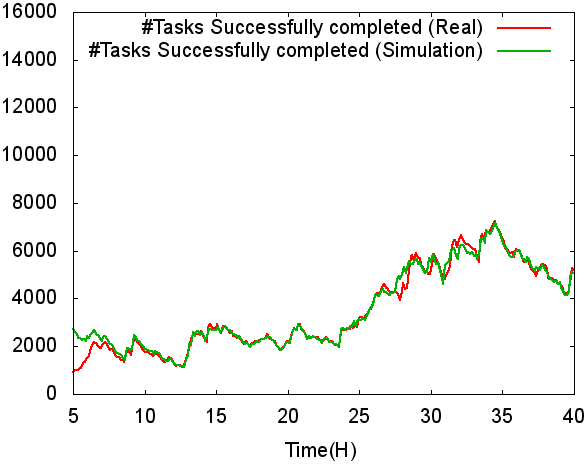}\label{fig:completed}} 
    \qquad
   \subfloat[Number of tasks waiting]{  \includegraphics[width=5.49cm]{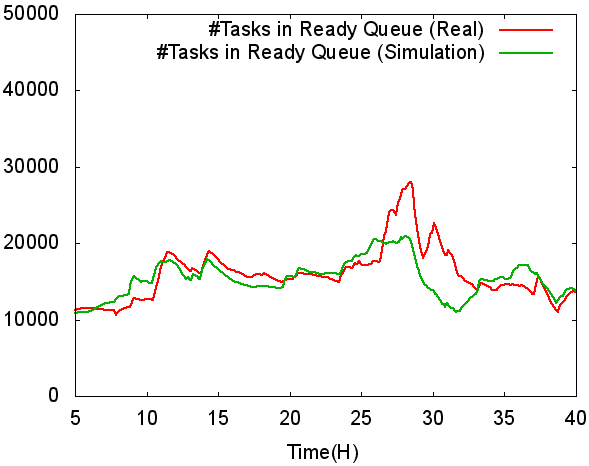}\label{fig:waiting}} 
    \qquad
   \subfloat[Number of tasks evicted]{  \includegraphics[width=5.49cm]{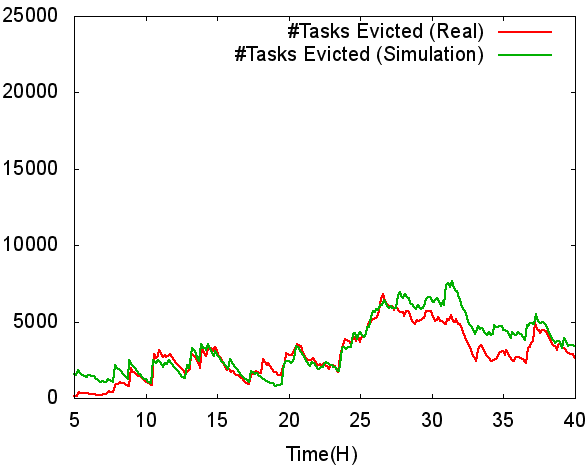}\label{fig:evicted}} 
\caption{Simulation and real data for four different metrics. All show good agreement between the behavior of our model and that of the real system.}      
\end{figure}

To assess the accuracy of the simulation results we perform a transient analysis, comparing the output of the simulator with the real data from the traces. Specifically, four metrics were considered: number of running tasks (Figure~\ref{fig:running}), number of completed tasks (Figure~\ref{fig:completed}), number of waiting tasks (ready queue size, Figure~\ref{fig:waiting}) and number of evicted tasks (Figure~\ref{fig:evicted}). Comparison of the real and simulated values can bring important evidence whether the model is able to reproduce the behavior of the real Google cluster. 

\begin{table}[b]
\begin{tabular}{|p{5.5cm} |p{1.15cm} |p{1.45cm} |p{1cm} |p{1cm}| }
\hline%\noalign{\smallskip}
Evaluation criterion&Running tasks&Completed tasks & Waiting tasks& Evicted tasks   \\
%\noalign{\smallskip}
\hline\hline
%\noalign{\smallskip}
Mean value obtained from the simulation&134476&3671.3&
15400.6&
3671.32\\
Mean value shown in the real traces&
136152&
3654.6&
15893.9&
2895.76\\
Maximum error (absolute value)&
4622&
1974&
9318&
2639\\
Maximum error (in percentage w.r.t. the mean value)&
3.40\%&
56.00\%&
59.00\%&
92\%\\
Mean error (absolute value)&
1858&
246&
1944&
755\\
Mean error (in percentage w.r.t. the mean value)&
0.01\%&
7.00\%&
12.20\%&
26\%\\
%\noalign{\smallskip}
\hline
\end{tabular}
\caption{Statistics of four evaluation criteria at intervals of 450 seconds.}
\label{tab:stat}     
\end{table}

All plots show the time series extracted from the trace data (green lines) and those produced by our model (red lines), with the additional application of exponential smoothing (to both) to reduce fluctuations. The figures show a very good agreement between the simulation results and the actual data from the traces. This means that the model provides a good approximation of the Google cluster.

We executed ten simulation runs; due to the fact that the model is deterministic (the only variation is in the choice of the machine where to execute a certain process), there are small differences across the runs. We report in Table~\ref{tab:stat} several statistics regarding the running, completed, waiting and evicted tasks.  These results are collected at intervals of 450 seconds.

%%%%%%%%%%%%%%%%%%%%%%%%%%%%%%%%%%%%%%%%%
%%%%%%%%%%%%%%%%%%%%%%%%%%%%%%%%%%%%%%%%%
%%%%%%%%%%%%%%%%%%%%%%%%%%%%%%%%%%%%%%%%%
\section{Related work}
\label{sec:discussion}

With the public availability of the two cluster traces~\cite{googleData} generated by the Borg system at Google~\cite{verma2015}, numerous analyses of different aspects of the data have been reported. These provide general statistics about the workload and node state for such clusters ~\cite{Liu2012,Reiss2012,Reiss2012a} and identify high levels of heterogeneity and dynamicity of the system, especially in comparison to grid workloads ~\cite{Di2012}. Heterogeneity at user level -- large variations between workload submitted by the different users  -- is also observed~\cite{abdul2014}. Prediction is attempted for job~\cite{guan2013} and machine~\cite{sirbu2015} failures and also for host load~\cite{di2012host}. However, no unified tool for studying the different traces were introduced. \emph{BiDAl} is one of the first such tools facilitating Big Data analysis of trace data, which underlines similar properties of the public Google traces as the previous studies. Other traces have been analyzed in the past~\cite{Kavulya2010,Chen2011,Chen2012}, but again without a general-purpose tool available for further study.

\emph{BiDAl} can be very useful in generating synthetic trace data. In general synthesizing traces involves two phases: characterizing the process by analyzing historical data and generation of new data. The aforementioned Google traces and log data from other sources have been successfully used for workload characterization. In terms of resource usage, classes of jobs and their prevalence can be used to characterize workloads and generate new ones~\cite{Mishra2010,Wang2011}, or real usage patterns can be replaced by the average utilization~\cite{Zhang2011}. Placement constraints have also been synthesized using clustering for characterization~\cite{Sharma2011}. Our tool enables workload and cloud structure characterization through fitting of distributions that can be further used for trace synthesis. The analysis is not restricted to one particular aspect, but the flexibility of our tool allows the the user to decide what phenomenon to characterize and then simulate.

Traces (either synthetic or the exact events) can be used for validation of various workload management algorithms. The Google trace has been used recently in~\cite{iglesias2014} to evaluate consolidation strategies, in~\cite{breitgand2014,caglar2014} to validate over-committing (overbooking), in~\cite{zhang2014dynamic} to perform provisioning for heterogeneous systems and in~\cite{di2013optimization} to investigate checkpointing algorithms. Again, data analysis is performed individually by the research groups and no specific tool was published. \emph{BiDAl} is very suitable for extending these analyses to synthetic traces, to evaluate algorithms beyond the exact timeline of the Google dataset.

Recently, the Failure Trace Archive (FTA) has published a toolkit for analysis of failure trace data~\cite{Javadi2013}. This toolkit is implemented in Matlab and enables analysis of traces from the FTA repository, which consists of about 20 public traces. It is, to our knowledge, the only other tool for large scale trace data analysis. However, the analysis is only possible if traces are stored in the FTA format in a relational database, and is only available for traces containing failure information. \emph{BiDAl} on the other hand provides two different storage options, including HDFS, with transfer among them transparent to the user, and is available for any trace data, regardless of what process it describes. Additionally, usage of FTA on new data requires publication of the data in their repository, while \emph{BiDAl} can be used also for sensitive data that cannot be made public. 

Although public tools for analysis of general trace data are scarce, several large corporations reported to have built in-house custom applications for analysis of logs. These are, in general, used for live monitoring of the system, and analyze in real time large amounts of data to provide visualization that help operators make administrative decisions. While Facebook use Scuba~\cite{Abraham2013}, mentioned before, Microsoft have developed the Autopilot system~\cite{Isard2007}, which helps with the administration of their clusters. Autopilot has a component (Cockpit) that analyzes logs and provides real time statistics to operators. An example from Google is CPI2~\cite{Hagmann2013} which monitors Cycles per Instruction (CPI) for running tasks to determine job performance interference; this helps in deciding task migration or throttling to maintain high performance of production jobs. All these tools are, however, not open, apply only to data of the corresponding company and sometimes require very large computational resources (e.g., Scuba). Our aim in this paper is to provide an open research tool that can be used also by smaller research groups that have more limited resources.

In terms of simulation, numerous modeling tools for computer systems have been introduced, ranging from queuing models to agent-based and other statistical models. The systems modeled range from clusters to grids, and more recently, to clouds and data centers~\cite{Zhao2012}. CloudSim is a recent discrete event simulator that allows simulation of virtualized environments~\cite{Calheiros2011}. More specialized simulators such as MRPerf have been designed for MapReduce environments~\cite{Wang2009}. In general, these simulators are used to analyze the behavior of different workload processing algorithms (e.g., schedulers) and different networking infrastructures. A comprehensive model is GDCSim (Green Data Centre Simulator), a very detailed simulator that takes into account computing equipment and its layout, data center physical structure (such as raised floors), resource management and cooling strategies~\cite{Gupta2011}. However the level of detail limits scalability of the system. Our simulator is more similar to the former examples and allows for large scale simulations of workload management (experiments with 12k nodes).

%%%%%%%%%%%%%%%%%%%%%%%%%%%%%%%%%%%%%%%%%
%%%%%%%%%%%%%%%%%%%%%%%%%%%%%%%%%%%%%%%%%
%%%%%%%%%%%%%%%%%%%%%%%%%%%%%%%%%%%%%%%%%
\section{Conclusions}
\label{sec:conclusion}

In this paper we presented \emph{BiDAl}, a framework that facilitates use of Big Data tools and techniques for analyzing large cluster traces. We discussed  a case study where we successfully applied \emph{BiDAl} to analyze Google trace data in order to derive workload parameters required by an event-based model of the cluster. Based on a modular architecture, \emph{BiDAl} currently supports two storage backends based on SQlite and Hadoop, while other backends can be easily added. It uses a subset of SQL as a common query language that is automatically translated to the appropriate commands supported by each backend. Additionally, data analysis using R and Hadoop MapReduce is possible. 

Analysis of the Google trace data consisted of extracting distributions of several relevant quantities, such as number of tasks per job, resource consumption by tasks, etc. These parameters were easily computed using our tool, showing how this facilitates Big Data analysis even to users less familiar with R or Hadoop. 

The model was analyzed under two scenarios. In the first scenario we performed a real-trace-driven simulation, where the input data were taken directly from the real traces. The results produced by the simulation in this scenario are in good agreement with the real data. The fidelity was obtained by fine tuning the model in terms of available resources, which accounts for unknown policies in the real cluster. Our analysis showed that reducing available memory to 48.9\% produces a good estimate of the actual data. In the second scenario we used \emph{BiDAl} to produce synthetic inputs by fitting the real data to derive their distribution. In this scenario the average values of the output parameters are in good agreement with the average values observed in the traces; however, the general shape of the output distributions are quite different. These differences could be due to over-simplifications of the model, such as the fact that only average values for resource consumption are used, or that the task arrival process is not modeled accurately.  Improvements of the accuracy of the model will be the subject of future work.

At the moment, \emph{BiDAl} can be used for pre-processing and initial data exploration; however, in the future we plan to add new commands to support machine learning tools for predicting abnormal behavior from log data. This could provide new steps towards achieving self-* properties for large scale computing infrastructures in the spirit of Autonomic Computing.

In its current implementation, \emph{BiDAl} is useful for batch analysis of historical log data, which is important for modeling and initial training of machine learning algorithms. However, live log data analysis is also of interest, so we are investigating the addition of an interface to streaming data sources to our platform. Future work also includes implementation of other storage systems, especially to include non-relational models. Improvement of the GUI and general user experience will also be pursued.

%\begin{acknowledgements}
%If you'd like to thank anyone, place your comments here
%and remove the percent signs.
%\end{acknowledgements}

% BibTeX users please use one of
%\bibliographystyle{spbasic}      % basic style, author-year citations
\bibliographystyle{spmpsci}      % mathematics and physical sciences
%\bibliographystyle{spphys}       % APS-like style for physics
%\bibliography{}   % name your BibTeX data base

\bibliography{refs,cloud}

\end{document}